\documentclass[aps,prl,preprint,superscriptaddress]{revtex4}

\usepackage{graphicx}
\usepackage{amsmath}
\usepackage{bm}
\usepackage{multirow}

\begin{document}


\title{Precise Numerical Solutions of Potential Problems
Using Crank-Nicholson Method}


\author{Daekyoung Kang}
\affiliation{Department of Physics, Korea University, Seoul 136-713, Korea}
\author{E. Won}
\affiliation{Department of Physics, Korea University, Seoul 136-713, Korea}


\date{\today}

\begin{abstract}
A new numerical treatment in the Crank-Nicholson method
with the imaginary time evolution operator is presented
in order to solve the Schr\"{o}dinger
equation.
The original time evolution technique is extended to a new operator that
provides a systematic way to calculate not only eigenvalues of ground state
but also of excited states.
This new method systematically produces 
eigenvalues with accuracies of
eleven digits with the Cornell potential that covers non-perturbative
regime. 
An absolute error estimation technique based on a power counting
rule is implemented.  This 
method is examined with exactly solvable problems and produces the
numerical accuracy down to 10$^{-11}$. 
\end{abstract}


\maketitle


Numerical computation of 
analytically unsolvable Schr\"{o}dinger equations has been of 
interest in atomic and molecular physics, 
quantum chromodynamics, and Bose-Einstein condensation of
trapped atomic vapors~\cite{Mehta,Vrscay,Chhajlany,Chiofalo}.
Conventionally, a wave function has been
represented as a linear combination of plane waves or of atomic 
orbitals~\cite{cayley}.
However, these representations entail high computational cost to
calculate the matrix elements for these bases. The plane wave bases
set is not suitable for localized orbitals, and the atomic orbital 
bases set is not suitable for spreading waves. In particular, potential
problems such as the Cornell potential~\cite{Eichten:1978tg,Eichten:1979ms}
are difficult to compute precise eigenvalues 
because they have singularities at the origin and at the infinity, and includes
non-perturbative regime when the linear term is significant. 

To overcome these problems, numerical methods such as
Land\'e subtraction~\cite{Norbury} and Nystrom plus correction~\cite{Tang} in 
the momentum space produced eigenvalues with six or seven digits.
Others adopted real-space
representation~\cite{Jacobs:1986gv,Roychoudhury}. 
In these methods, a wave function is discretized by
grid points in real space providing from five to seven digits. Also,
estimates using the exact solutions of the Killingbeck
potential produced eigenvalues with seven digits~\cite{Chhajlany}.
Among these real-space methods, a method called
Crank-Nicholson (C-N) scheme~\cite{recipes,galbrath}
is known to be especially useful for one-dimensional
systems because this method conserves the norm of the wave function exactly
and the computation is stable and accurate even in a long time
slice. These characteristics are very attractive for solving the
Cornell potential 
in order to compute eigenvalues and eigenstates of the
system precisely.

The current numerical precisions in relevant research areas may not
be as high as one would hope to achieve. For example,
it may forbid one to study fine or hyper-fine
structure effects in the atomic system~\cite{Norbury}. 
Calculation of matrix elements subject to large subtraction may
require high accuracy
in quantum chromodynamics~\cite{Bodwin:2006dn,Bodwin:2006dm}. 
Also, none of the references we compiled for this
study contains the
serious error estimate on their numerical calculations, which is an
important indicator
of the reliability of a suggested numerical method.

In this letter, we apply the C-N method to solve the 
Schr\"{o}dinger equation
and present two new numerical methods. 
First,
the C-N
method with the imaginary time evolution operator is re-interpreted
by extending its allowed region. This
method produces ground-state eigenvalues with numerical accuracies
of eleven digits when the Cornell potential is used. 
We then extend the original time evolution technique to a new operator that
provides a systematic way to calculate not only 
the eigenvalue of the ground state
but also those of excited states with less computational time by a
factor of 10, 
while providing
the same accuracy as in our re-interpreted C-N method. 
Our methods will be useful in calculations of 
eigenstates 
for the Cornell potential
and we discuss some of the results.
At the end, we discuss a mathematically simple but rigorous 
absolute error estimation on the
numerical calculations presented in this letter.

 C-N method~\cite{recipes,galbrath} is
a finite difference method used for solving diffusive
initial-value problems numerically.
The main part of this method is the time evolution operation 
and the evolution operator for the Schr\"{o}dinger
equation may be approximated by Cayley's form~\cite{cayley} 
as
\begin{equation}\label{eq:evolution1}
e^{-i\mathcal{H} t} =
\frac{1-\frac{i}{2}\mathcal{H} t}{1+\frac{i}{2}\mathcal{H} t}
+ \mathcal{O} (\mathcal{H}^3 t^3),
\end{equation}
where $\mathcal{H}$ is the Hamiltonian of the problem.
This equality is correct up to the second order in $\mathcal{H} t$ and the
approximation is valid when $|\mathcal{H} t|$ $\ll$ 1.
By multiplying this operator to an initial wave function, one obtains 
the evolved wave 
function. 
The standard C-N method makes use of Eq.~(\ref{eq:evolution1}) in order
to study the time evolution of the wave function~\cite{recipes,galbrath}. 

 Next, we introduce the imaginary time method~\cite{Chiofalo}
to calculate the eigenfunctions and eigenvalues.
By a Wick rotation, 
one replaces $t$ by $-i\tau$ in the time evolution
operator~\cite{Chiofalo}. This transforms the 
original Schr\"{o}dinger equation into a diffusion
equation. 
Then a wave function evolves in time slice $\Delta t$ as 
\begin{equation}\label{eq:img}
u(\rho,\Delta t)=\sum_{i=1}^{\infty} C_i u_i(\rho) e^{-i\zeta_i \Delta t}
=\sum_{i=1}^{\infty} C_i u_i(\rho) e^{-\zeta_i\Delta\tau},
\end{equation}
where $u_i(\rho)$ and $\zeta_i$ are the eigenfunction and the eigenvalue 
for $i$-th state, respectively. $C_i$ is the relative amplitude for 
$i$-th state 
and the summation is over all possible eigenstates of the system.

For the imaginary-time version, eigenfunctions decay monotonically in time 
till the steady state is reached. The ground-state 
eigenvalue can then be read off from the steady-state eigenfunction 
as $\tau$ $\rightarrow$ $\infty$~\cite{Chiofalo}.
Therefore, we acknowledge that
the time evolution operation itself in the C-N method acts as a tool
that selects the ground state only. 
Here, we advocate 
that the condition $|\mathcal{H} \tau| \ll$ 1 is not necessary in the
calculation of the ground-state eigenvalue.
When all the eigenvalues are negative such as the pure Coulomb potential
case, where 
$\zeta_0 <\cdots<\zeta_n<\cdots<0$, the amplification of the
ground-state coefficient may happen 
in the region $-2<\mathcal{H}\tau<0$ as the time evolution continues.
On the other hand, when all the eigenvalues are positive 
such as in problems with
the linear and Cornell potentials, where 
$0<\zeta_0<\cdots<\zeta_{n}$,
the time evolution operator 
can amplify the ground-state coefficient in the region
$\mathcal{H\tau}<-2$.
Note that
we relax the condition $|\mathcal{H} \tau| \ll 1$ in order to obtain
faster convergence to ground-state eigenfunction 
even if that region is used in the standard C-N method.
We call this approach as a {\it relaxed C-N method}. 

Once the ground-state wave function is obtained, one can obtain the 
ground-state
eigenvalue from the expectation value of the Hamiltonian of interest.
Note that, for the numerical computation of an expectation value,
the upper bound of the integral can not be an infinity but a cut-off 
value, $\rho_{\textrm{max}}$. We will
explain how to control the numerical error produced by ignoring the region 
$(\rho_{\textrm{max}},\infty)$ later.

 The spherical Stark effect in hydrogen and 
a bound state for a heavy quarkonium may be described by 
the Schr\"{o}dinger equation
with the Cornell potential~\cite{Eichten:1978tg,Eichten:1979ms,Vrscay}.
The radial part of the Schr\"{o}dinger equation 
with orbital-angular-momentum quantum number $\ell$ that is 
relevant to such problems may be given by
\begin{equation} \label{eq:schrodinger}
\left[
-\frac{d^2}{d\rho^2}+\frac{\ell(\ell+1)}{\rho^2}-\frac{\lambda}{\rho}+\rho
\right] u(\rho)=\zeta u(\rho), 
\end{equation}
where the dimensionless wave function $u(\rho)$ and the dimensionless energy
eigenvalue $\zeta$ in Eq. (\ref{eq:schrodinger}) 
are described in Ref.~\cite{Eichten:1978tg}. The parameter
$\lambda$ is the relative strength between the Coulomb and the linear
potentials. We call it as the coulombic parameter throughout this letter.

 The relaxed C-N method is now applied to the Cornell potential 
problem.
The
$\lambda$ dependence of the ground-state eigenvalues 
in Ref.~\cite{Eichten:1978tg} is reproduced with our relaxed C-N
method and a comparison of two results is
summarized in Table~\ref{table:eigen1}. 
The time evolution in our relaxed C-N method
gives eigenvalues over iterations
giving stable sixteen-digits and all agree well with
the results in Ref.~\cite{Eichten:1978tg}. 
We also find that the convergence speed is improved
by 10 times compared with the standard C-N
method that we tested. For excited states, one can in principle
obtain the eigenvalues from the lowest to higher
states by the Gram-Schmidt orthogonalization procedure. We find that the
standard C-N method is not useful in calculating
excited states because the convergence speed is practically zero when
we require the numerical precision to be 10 digits. 
However, with the relaxed C-N method, 
the excited-state eigenvalues are successfully found, with the
speed being 10 times
slower than the time needed to find the ground state.
The amount of time required to calculate excited states is no longer than
ten minutes with a Pentium IV CPU with 0.5 GB memory.
\begin{table}
\begin{tabular}{c|c|c|c}
\hline
\hline
$\lambda$&$\zeta$(Ref.\cite{Eichten:1978tg})
&$\zeta$ (this work) &$\Delta\zeta$\\
\hline
0.0 ~~~&2.338~107&~2.338~107~410~458~750~~&$1.0\times10^{-12}$\\
0.2 ~~~&2.167~316&~2.167~316~208~771~731~~&$1.0\times10^{-12}$\\
0.4 ~~~&1.988~504&~1.988~503~899~749~943~~&$9.6\times10^{-13}$\\
0.6 ~~~&1.801~074&~1.801~073~805~646~145~~&$8.5\times10^{-13}$\\
0.8 ~~~&1.604~410&~1.604~408~543~235~973~~&$6.6\times10^{-13}$\\
1.0 ~~~&1.397~877&~1.397~875~641~659~578~~&$3.8\times10^{-13}$\\
1.2 ~~~&1.180~836&~1.180~833~939~744~863~~&$2.1\times10^{-14}$\\
1.4 ~~~&0.952~644&~0.952~640~495~219~193~~&$5.8\times10^{-13}$\\
1.6 ~~~&0.712~662&~0.712~657~680~462~421~~&$1.3\times10^{-12}$\\
1.8 ~~~&0.460~266&~0.460~260~113~875~977~~&$2.3\times10^{-12}$
\\
\hline
\hline
\end{tabular}
\caption{\label{table:eigen1}
Dependence of the ground-state 
eigenvalues $\zeta$ on the coulombic parameter $\lambda$. The Cornell
potential is used in the calculation.
The second column 
contains the numerical results from Ref.~\cite{Eichten:1978tg}
and the third column contains our result with the relaxed C-N
method.
The number of grid points is set to be $N$ = 300,000 and $\Delta \tau$
= $-$ 5 for the numerical analysis.
An estimate on the numerical errors 
of the computation $\Delta \zeta$ 
is also listed at the last column and will be discussed later.}
\end{table}

 We emphasize that the condition 
$|\mathcal{H}\tau| \ll$ 1 
can be relaxed as far as the goal of the numerical computation is 
to obtain the ground-state eigenfunction. We extend this idea further
in order to calculate excited-state eigenvalues and eigenfunctions
systematically and more efficiently.

Consider the following operator
\begin{equation}\label{eq:amp}
\frac{1}{\mathcal{H}-\beta},
\end{equation}
where $\beta$ is an arbitrary real number in the eigenvalue space.
If we apply the operator in Eq.~(\ref{eq:amp}) to the wave packet $k$ times,
\begin{equation}
\left( \frac{1}{\mathcal{H}-\beta}\right)^k u(\rho)
=\sum_{i=1}^{\infty}
\frac{C_{i}}{(\zeta_i -\beta)^k} u_i(\rho),
\end{equation}
where we assume that the 
time independent wave packet $u(\rho)$ can be expressed as a linear 
combination of
the infinite number of eigenfunctions $u_i(\rho)$ with coefficients $C_i$.
For $\beta=\zeta_i +\epsilon$ such that $|\epsilon| \ll 1$, we have 
\begin{equation}
\left| \frac{1}{\zeta_i-\beta} \right| \gg
\left| \frac{1}{\zeta_j-\beta} \right|
\end{equation}
if $j\neq i$. Therefore, Eq.~(\ref{eq:amp}) plays a role as an amplifier 
which magnifies the contributions of the 
term with the nearest eigenvalue from the point $\beta$. 
In this way, all the eigenvalues within an arbitrary range in $\beta$ can
be found systematically, by running $\beta$ within the range.
We call this approach as a {\it modified C-N method}.
Advantages of this method are as follows. First, the calculations of
excited states can be carried out systematically by stepping
through different values of $\beta$. Second,
the computing time for 
calculating excited-state wave functions or eigenvalues
is similar to that needed for the ground
state. 
Furthermore, it does not lose accuracies in the calculation of higher-state
eigenvalues while other methods do often.
This contrasts the modified C-N method to the
relaxed C-N method. 
With the relaxed C-N method, the Gram-Schumidt
orthonormalization slows down the computing speed by ten times, as
mentioned before.

 First, the modified C-N method is tested on 
the pure Coulomb potential 
and to the pure linear potential ($\lambda$ = $0$) where the exact eigenvalues
are known for both cases. This is a good benchmark 
because one directly examines
the performance of the algorithm by comparing the exact solutions to the
numerical results of the algorithm to be tested. 
We find that the numerical values of eigenvalues 
agree well with known analytical values, up to eleven digits.
Second, the modified C-N method is applied to the Cornell potential
and reproduce the result in 
Table~\ref{table:eigen1}. We find that 
eleven digits of the eigenvalues are reproduced completely
from $\lambda$ = 0.0 to 1.8. 
This is consistent
with our error estimation which will be explained later.
Third, 
the modified C-N method 
is applied to the calculation of excited states.
Table~\ref{table:eigen2} shows the eigenvalues obtained. Note that
for the $1S$ state in Table~\ref{table:eigen2},
one can compare the eigenvalue with the number in Table~\ref{table:eigen1}
and only the last two digits are different, which is again consistent with
our error estimation. 
Again, for the ground state, the relaxed and the
modified C-N methods require similar amount of computing time,
but for the excited states, the modified C-N method is
faster by 10 times.
We checked the computing speed for the Coulomb, linear, and Cornell
potentials separately and all three give similar performances.
\begin{table}
\begin{tabular}{c|c|c}
\hline
\hline
State&$\zeta$&$\Delta\zeta$\\
\hline
$1S$ &1.397~875~641~659~581 &$3.8\times 10^{-13}$\\
$2S$ &3.475~086~545~392~783 &$3.4\times 10^{-12}$\\
$3S$ &5.032~914~359~529~781 &$6.3\times 10^{-12}$\\
$4S$ &6.370~149~125~476~954 &$9.4\times 10^{-12}$\\
$5S$ &7.574~932~640~578~566 &$1.3\times 10^{-11}$\\
$1P$ &2.825~646~640~702~388 &$1.2\times 10^{-12}$\\
$2P$ &4.461~863~593~453~813 &$3.1\times 10^{-12}$\\
$3P$ &5.847~634~227~299~904 &$5.5\times 10^{-12}$\\
$1D$ &3.850~580~006~802~002 &$5.9\times 10^{-13}$\\
$2D$ &5.292~984~139~140~243 &$2.0\times 10^{-12}$
\\
\hline
\hline
\end{tabular}
\caption{\label{table:eigen2} 
Numerical values of eigenvalues and the error estimation 
for various excited states with the Cornell potential.
The modified C-N method is used to calculate eigenvalues.
The coulombic parameter is set to $\lambda=1.0$ and the number of 
grid points to be $N=300,000$.}
\end{table}

 There are two sources of errors in our numerical calculation.
One is the 
cut-off ($\rho_{\textrm{max}}$) and the other is the discretization of 
continuous equations.
The cut-off gives the imperfect numerical 
integration but could in principle be reduced as small as one wishes 
by increasing the value of  
$\rho_{\textrm{max}}$. We estimate the error due to 
the finite value of $\rho_{\textrm{max}}$ for the Coulomb and 
linear potentials, respectively, by integrating 
exact eigenfunctions. We find that the error 
is 10$^{-15}$ or smaller when $\rho_{\textrm{max}}$ = 20, for example. 
In fact, for the practical purpose,
we control the value of $\rho_{\textrm{max}}$ in such a way that
the numerical error due to finite value of $\rho_{\textrm{max}}$ 
is smaller than the error from  
the fact that we deal with discrete processes. 
This assumes that the error estimates
with the Coulomb and linear potentials individually are 
not significantly different
from the errors due to the Cornell potential. Practically,
selecting proper values of 
$\rho_{\textrm{max}}$ 
is important, because too small values cause large errors and 
too large values may slow down the computation. We find that
$\rho_{\textrm{max}}$ = 30 is an optimal value for most of our
applications shown in this letter. Note that $\rho_{\textrm{max}}$ = 30
corresponds to 30 times of the Bohr radius in the hydrogen atom
problem, for example.
 
More serious source of the error is originated from the discretization of
continuous differential equations.
In this letter, differentiation and integration of wave functions
are discretized with the following prescriptions~\cite{recipes}
\begin{subequations}
\begin{eqnarray}
u''_{j}
&=&\frac{u_{j+1} -2u_{j}+u_{j-1}}{\Delta\rho^2}+\mathcal{O}(\Delta\rho^2),
\label{eq:diff2} \\
\int d\rho\, u(\rho)&=&\frac{\Delta\rho}{2}
					\sum_{j=1}^{N}
					\left( u_{j+1}+u_{j}\right)
					+\mathcal{O}(\Delta\rho^2),
\label{eq:int} 
\end{eqnarray}
\end{subequations}
where $\Delta \rho$ is the distance between two nearest discrete points 
in the calculation.  
\begin{figure}
\includegraphics[width=17.0cm]{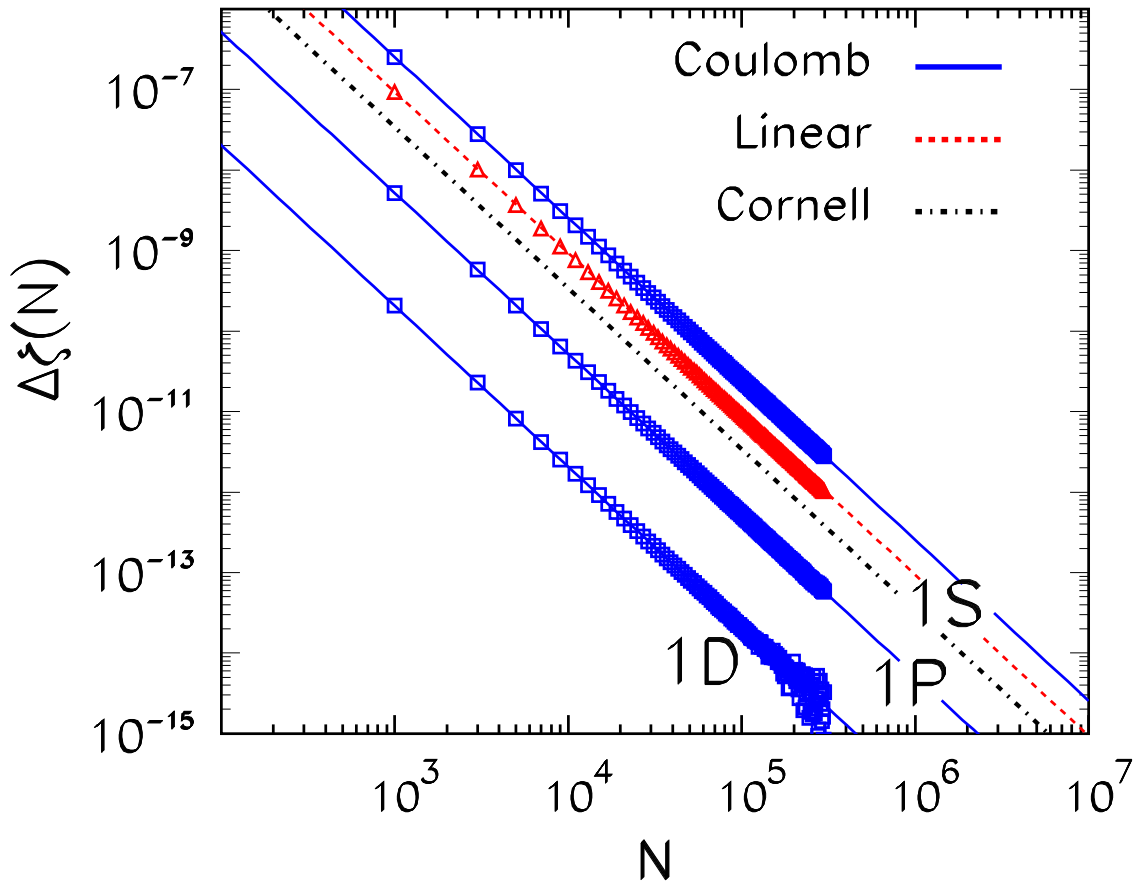}
\caption{Numerical error estimation for the eigenvalues
as a function of the number of
grid points. For the Coulomb (linear) potential, 
$|\zeta_{\textrm{exact}}$ $-$
$\zeta(N)|$ is indicated as open boxes (triangles) for $1S$, $1P$, 
and $1D$ states
($1S$ state only for the linear potential).
The error estimates based on Eq.~(\ref{eq:estimation}) for
the Coulomb, linear, and Cornell potentials are shown in
straight, dashed, and dot-dashed lines, respectively.
Note that both axes are in logarithmic scales.}
\label{fig:error}
\end{figure}
From both Eqs.~({\ref{eq:diff2}) and (\ref{eq:int}),
numerical errors contained in the discretization are proportional to 
$\Delta\rho^2=N^{-2}$ where $N$ is the number of grid points in the
discretization. Therefore, the error in eigenvalues due to the discretization 
may be approximated as
\begin{equation}\label{eq:estimation}
\Delta\zeta(N)\approx c N^{-2},
\end{equation}
where $\Delta\zeta(N)=|\zeta_{\textrm{exact}}-\zeta(N)|$ and the constant 
$c$ depends on the potential.
If we select two arbitrary values in grid points,
$N_1$ and $N_2$, for example, then we can easily obtain the constant $c$ as
\begin{equation}\label{eq:a}
c\approx
\left| \frac{\Delta\zeta(N_1)-\Delta\zeta(N_2)}{N_1^{-2}-N_2^{-2}}
\right|
=
\left| \frac{\zeta(N_1) - \zeta(N_2)}{N_1^{-2}-N_2^{-2}}
\right|.
\end{equation}
With this, one can estimate the error due to the discretization prescription
in Eqs.~({\ref{eq:diff2}) and (\ref{eq:int}).
We refer it as an error estimation from the power counting rule. 
In Fig.~\ref{fig:error}, our estimate of the numerical error 
$\Delta\zeta(N)$
is compared with the true error $|\zeta_{\textrm{exact}}$ $-$ $\zeta(N)|$
for the Coulomb and linear potentials. 
Here we used our modified C-N method for the error analysis.
It is apparent that our error
estimate is accurate down to $10^{-11}$ for the $1S$ state under 
the Coulomb potential, for example, when $N$=300,000. 
For others, the results are better as in Fig.~\ref{fig:error}.
For the $1D$ state under the Coulomb potential, 
the true error looks unstable at large value of $N$ and
we find that this is due to the limitation in storing significant digits 
during our computation of $\zeta(N)$.
Note that for the Cornell potential, the true errors cannot be calculated
because the exact solutions are unknown.
Therefore, for the Cornell potential, we estimate that the errors are 
in the range of $10^{-12}$ as in Fig.~\ref{fig:error}.
We use this error estimation technique throughout this letter and numerical
values of the error estimates are included in 
Tables~{\ref{table:eigen1} and \ref{table:eigen2}.

 In conclusion, we have presented two numerical methods for calculating
Schr\"{o}dinger equation with the Crank-Nicholson method. In the 
relaxed C-N method, the time evolution operator is re-interpreted as a
weighting operator for finding
the ground state eigenfunction more precisely. This idea is extended to
a new operator in the modified C-N method  
that is more efficient in computing not only  
the ground-state but also excited-state wave functions systematically. 
An absolute error estimation method is
presented based on a power counting rule and is consistent with 
predictions when exact solutions are known.  
These two algorithms may be useful when precise numerical
results are required. Possible applications may include 
Cornell potential~\cite{Eichten:1978tg,Eichten:1979ms}
and Bose-Einstein condensation of trapped atomic vapors~\cite{Chiofalo}.   

We thank Jungil Lee for his suggestion on this topic and 
Q-Han Park and Ki-hwan Kim for useful discussion on the numerical
treatment.
E.~W. is indebted to Tai Hyun Yoon for his critical comments on this
manuscript.
D.~K.'s research was supported
in part by the Seoul Science Fellowship of Seoul Metropolitan
Government
and
by the Korea Research Foundation Grant funded
by the Korean Government (MOEHRD), (KRF-2006-612-C00003).
E.~W.'s research was supported by grant No.
R01-2005-000-10089-0
from the Basic Research Program of the Korea Science
\& Engineering Foundation.



\newpage

\end{document}